\documentclass[useAMS,usenatbib,usegraphicx]{mn2e}

   \newcommand{\aap}{A\&A}
\newcommand{\aj}{AJ}         \newcommand{\apj}{ApJ}
\newcommand{\apjl}{ApJ}      \newcommand{\apjs}{ApJS}
\newcommand{\mnras}{MNRAS}   \newcommand{\nat}{Nature}
     \newcommand{\pasp}{PASP}

\title[Are galaxies with AGN a transition population?]
	{Are galaxies with AGN a transition population?}
\author[P. B. Westoby et al.]
	{P. B. Westoby\thanks{E-mail:pbw@astro.livjm.ac.uk}, C. G. Mundell and I. K. Baldry\\
	Astrophysics Research Institute, Liverpool John Moores University, Twelve Quays House, 
        Egerton Wharf, Birkenhead, CH41 1LD, UK}

\begin{document}

\date{Accepted 2007 September 22.  Received 2007 August 28; 
  in original form 2007 June 28}

\pagerange{\pageref{firstpage}--\pageref{lastpage}} \pubyear{2007}

\maketitle

\label{firstpage}

\begin{abstract}

We present the results of an analysis of a well-selected sample of galaxies with active and inactive
galactic nuclei from the Sloan Digital Sky Survey, in the range $0.01 < z < 0.16$. The SDSS galaxy
catalogue was split into two classes of active galaxies, Type~2 AGN and composites, and one set of
inactive, star-forming/passive galaxies. For each active galaxy, two inactive control galaxies were
selected by matching redshift, absolute magnitude, inclination,
and radius. The sample of inactive galaxies naturally divides into a red and a blue sequence, while
the vast majority of AGN hosts occur along the red sequence. In terms of H$\alpha$ equivalent width, the
population of composite galaxies peaks in the valley between the two modes, suggesting a transition
population. However, this effect is not observed in other properties such as colour-magnitude space,
or colour-concentration plane.  Active galaxies are seen to be generally bulge-dominated systems,
but with enhanced H$\alpha$ emission compared to inactive red-sequence galaxies.  AGN and composites
also occur in less dense environments than inactive red-sequence galaxies, implying that the fuelling 
of AGN is more restricted in high-density environments. These results are therefore inconsistent with
theories in which AGN host galaxies are a `transition' population.  We also introduce a systematic
3D spectroscopic imaging survey, to quantify and compare the gaseous and stellar kinematics of a
well-selected, distance-limited sample of up to 20 nearby Seyfert galaxies, and 20 inactive control
galaxies with well-matched optical properties. The survey aims to search for dynamical triggers of
nuclear activity and address outstanding controversies in optical/IR imaging surveys.

\end{abstract}


\section{Introduction} 

Active Galactic Nuclei (AGN) have long been considered a curiosity in their own right \citep{schmidt63,
lybell69,blan74}, but are now recognised to be integral to galaxy formation and evolution. 
At early cosmological epochs, gas-rich galaxies are thought to form and
collide in violent mergers \citep{mihos96}, triggering vast bursts of star formation, and fuelling
supermassive black holes in their cores \citep{kriek06}.  Recent simulations of isolated and merging
galaxies by \cite{spring05} have incorporated feedback from star formation and black-hole accretion,
and find that once an accreting supermassive black hole (SMBH) has grown to some critical size, the
AGN feedback terminates its growth as a large fraction of the remaining nuclear gas is driven out by 
the powerful quasar.

In the current epoch, the peak of the quasar era is over and the galaxy merger rate has declined
\citep{struck97}.  Nevertheless, at least 20\% of today's galaxies show scaled-down quasar activity
in their centres, and direct measurements of active and inactive galaxy dynamics have revealed a
tight correlation between the central black-hole mass and host galaxy stellar velocity dispersion or
bulge mass. This $M_{\bullet}-\sigma$ relation \citep{geb00,merr01} points to an intimate link
between host galaxy evolution and central black-hole growth, suggesting that all bulge-dominated
galaxies today harbour dead quasars and that a lack of nuclear activity cannot be attributed to the
absence of a central black hole.  Therefore, given the ubiquity of supermassive black holes, what
determines the degree of nuclear activity in todays galaxies and what role is played by the host
galaxy in triggering and fuelling their dormant black holes remain open issues.

Previously, studies of nearby active galaxies were based on small galaxy samples, like the C$f$A
Seyfert Sample \citep{huchra92}, the 12 Micron Active Galaxy Sample \citep{rush93} and the
\cite{ho95} galaxy sample of approximately 486 galaxies covering a range of activity types. Now,
however, the standard has been set by the Sloan Digital Sky Survey (SDSS), from which a range of
useful samples of both active and inactive galaxies can be selected in a well-defined, uniform way.

The SDSS \citep{york00} provides, for the first time, 5-band photometry and spectroscopy of many
thousands of low redshift AGN \citep{kauf03c,hao05a}, enabling constraints to be placed on galaxy and AGN
evolution over a wide range of galaxy masses, and acts as the definitive supporting data to all
detailed follow-up galaxy studies. Initial investigations into the growth and evolution of black
holes using these databases have yielded contrasting results --- the origin of the correlation
between galaxy bulge and central black-hole masses is hotly debated in particular.

\citet{heck04} find that the majority of present-day accretion occurs onto 10$^{8}$ solar-mass black
holes in moderate mass galaxies, suggesting that bulge and black-hole evolution is still tightly
coupled today, and that the evolution of AGN luminosity functions is driven by a decrease in the mass
scale of accreting black holes. In contrast, the \citet{hao05b} study of about 3\,000 SDSS AGN
concludes that evolution in AGN luminosity functions is driven by evolution in the Eddington ratio,
rather than black-hole mass. Meanwhile, \cite{grup04} and \cite{grup05}
use a sample of 75 X-ray selected AGN to argue that black holes in Narrow Line Seyfert 1 galaxies in
particular, grow by accretion in well-formed bulges to produce the $M_{\bullet}-\sigma$
relation over time, refuting theories for the origin of black-hole / bulge relations in which
black-hole mass is a constant fraction of bulge mass at all epochs or in which bulge growth is
controlled by AGN feedback \citep{king03,ferr02,kriek06}.
The subclass of NLS1s has been further explored more recently by \cite{komo07}, who find that 
NLS1s are accreting at a rate higher than the Eddington rate, confirming that their BHs must be growing. 
They suggest that either NLS1 galaxies evolve into Broad Line Seyfert 1 (BLS1) galaxies with respect to 
their black hole mass distribution, which would require some change in the bulge properties, possibly due to
feedback, or that NLS1s are just low-mass extensions of BLS1 galaxies, and the high accretion rate could just be 
caused by a relatively short-lived accretion phase.

Another characteristic revealed by large automated surveys such as the SDSS, is that the galaxy
population is found to be bimodal in colour \citep{lilly95,strat01}, with it now being more
natural to describe a galaxy as being on the ``red sequence'' or ``blue sequence'', rather than
being ``early type'' or ``late type'' \citep{bald06}.  A key goal of galaxy evolution theory is then
to explain the colour bimodality of galaxies, the relationships within each sequence, and where
active hosts fit into this picture. Therefore, the understanding of galaxy formation and evolution
processes necessitates full inclusion of AGN and their hosts.

Recently efforts have been made to try to understand where AGN host galaxies fall 
in colour-magnitude space. The red sequence consists of mainly massive, passively evolving galaxies,
while the majority of galaxies show blue colours, attributed to ongoing star formation. The two
sequences are separated by a relatively narrow valley in colour space.

The emerging consensus is that intense star formation is fuelled by galaxy mergers at high redshift,
which forms massive bulges, and at some point the star formation ceases, resulting in the galaxy
migrating from the blue sequence to the red sequence. AGN have been singled out as the mechanism for
star-formation quenching, leading to suggestions that AGN should occupy a distinct, `transition' 
region, of the colour-magnitude diagram (CMD). However this does not necessarily have to be the 
case, as the resulting galaxy could just be bluer due to increased star-burst activity, or redder 
due to enhanced dust (e.g., Luminous Infra-Red Galaxies), or some combination of the original 
colours of the merging galaxies. Young stellar populations and nuclear starbursts are therefore also 
important components in the dynamics of nuclear activty \citep{gonza98,sarzi07}.

\cite{nand07} studied the colour-magnitude relation for a sample of 50 X-ray selected AGN from
the AEGIS survey (All-wavelength Extended Groth strip International Survey), in the range $0.6 < z <
1.4$. They conclude that AGN fall on the red sequence or on the red edge of the blue sequence, with
many in between these two modes.

\cite{mart07} further explored the idea of a `transition' region by exploiting the separation of the 
blue and red sequences in the ${\rm UV} - r$ colour-magnitude diagram. Using a sample of UV selected 
galaxies from the GALEX Medium Imaging Survey, along with SDSS data, they explored the nature of 
galaxies in the transition zone. The AGN fraction of their sample was found to peak in the 
transition zone, and they also found circumstantial evidence that star-formation quenching 
rates were higher in higher luminosity AGN.

Higher image quality than the SDSS is provided by the Millennium Galaxy Catalogue (MGC; \citealt{liske03,cross04}), 
which has obtained redshifts for 10\,095 galaxies to $B < 20 {\rm\,mag}$, covering 37 deg$^{2}$ of
equatorial sky. The MGC study of galaxy bimodality was in the colour-structure plane,
(\citealt{drive06}; hereafter D06) and expressed a contrasting theory for the bimodality.  The MGC
survey has sufficient resolution to achieve reliable bulge-disc decomposition \citep{allen06}, and
in separating out the two components, D06 claim that galaxy bimodality is caused by the bulge-disc
nature of galaxies, and not by two distinct galaxy classes at different evolutionary stages. In this
case, they show that the bulge-dominated, early-type galaxies populate one peak and the bulge-less,
late-type galaxies occupy the second. The early- and mid-type spirals sprawl across and between the
peaks. They also propose that the reason for the dual structure of galaxies is due to galaxy
formation proceeding in two stages; first, there is an initial collapse phase, which forms the
centrally concentrated core and a black hole, and second, there is the formation of a planar
rotating disc caused by accretion of external material building up the galaxy disk.

In this paper, we present the properties of a magnitude-limited sample of ``active'' and
``inactive'' galaxies carefully selected from the SDSS, which also acts as the parent sample for
detailed followup studies using the IMACS-IFU on the Magellan 6.5m telescope. Our sample selection
is described in \S 3, the sample properties in \S 4, and implications of these properties in \S
5. We conclude by introducing the Magellan survey, which is now underway.

\section{The Sloan Digital Sky Survey}

The SDSS \citep{york00,stou02} is an imaging and spectroscopic survey, which, when completed, will
provide detailed optical images covering more than a quarter of the sky and a 3-dimensional map of
about a million galaxies and quasars. The survey uses a dedicated 2.5m telescope at Apache Point
Observatory, New Mexico, equipped with a large format mosaic CCD camera --- 30 CCDs in six columns
and five rows --- to image in five optical bands; $u$, $g$, $r$, $i$ and $z$, with effective 
wavelengths of 3\,550, 4\,670, 6\,160, 7\,470 and 8\,920 \AA\ \citep{fuku96,gunn98}.

The images are calibrated photometrically \citep{hogg01} and astrometrically \citep{pier03} using a
separate 0.5-m telescope, and reduced using a pipeline \textbf{\texttt{photo}} that measures the
observing conditions, and detects and measures objects. In particular, \textbf{\texttt{photo}}
produces various types of magnitude including `Petrosian', the summed flux in an aperture that
depends on the surface-brightness profile of the object. 
The magnitudes are Galactic extinction-corrected using the dust maps of
\cite{schl98}. Details of the imaging pipelines are given by \citet{stou02}. 

The main telescope is also equipped with two double fibre-fed spectrographs, covering a wavelength
range of 3800-9200 \AA, \,over 4\,098 pixels, with a resolution of $\lambda/\Delta\lambda \sim 1\,800$. 
Once a sufficiently large area of sky has been imaged, the data are analysed using
`targeting' software routines that determine the objects to be observed spectroscopically. The
targets are then assigned to plates, each with 640 fibres, using a tiling algorithm
\citep{blan03a}. The main restriction is that two fibres cannot be placed within 55\arcsec\ on the
same plate.

Spectra are typically taken using three 15-minute exposures in moderate conditions (the best
conditions are used for imaging). The signal-to-noise ratio ($S/N$) is typically 10 per pixel (1-2\AA)
for galaxies. The pipeline \textbf{\texttt{spec2d}} extracts, and flux- and wavelength-calibrates the
spectra.

\section{Sample Selection}

We start with the galaxy catalogue from the SDSS Fourth Data Release (DR4; \citealt{DR4}), as
distributed by the MPA Garching group.  The catalogue\footnote{Data publicly available at
http://www.mpa-garching.mpg.de/SDSS/DR4/} \citep{kauf03a,brinch04,trem04} contains photometric and
follow-up spectroscopic data for all objects spectroscopically classified as galaxies with Petrosian
$r$-band magnitudes less than 17.77.  The catalogue does not include objects classified as
`quasars' by the SDSS spectral classification algorithm. Broad emission-line galaxies (`Type~1'
Seyferts) are therefore generally excluded from the MPA catalogue, although there are still some
Type~1 Seyferts remaining if their spectra contain a significant fraction of stellar light. Galaxies
with a velocity dispersion, $\sigma$, of greater than 500 km s$^{-1}$ are therefore assumed to be
broadline Seyferts (Type~1 AGN), and due to the relatively low number of these (less than 4 per cent 
of the overall sample), are rejected from this analysis. In this paper therefore, we will only be 
comparing inactive galaxies with narrow-line galaxies (`Type~2' AGN, or `Type~2' Seyferts). Exclusion 
of type 1 AGN from our active sample does not affect the overall conclusions.

The MPA DR4 catalogue contains derived physical properties for 567\,486 galaxy spectra of 520\,738
individual galaxies, with duplicates being removed using a $1.5''$ matching radius. A redshift cut
of $0.01 < z < 0.16$ is then applied to all galaxies to limit evolution, k-corrections, and
low-redshift problems. 359\,154 DR4 galaxies then remain in our sample, from which active galaxies
are isolated, leaving most of the remaining galaxies to be used as potential control galaxies. Once
the AGN sample has been produced, two inactive control galaxies are selected for each AGN, closely
matched in their optical properties.

\subsection{AGN and Inactive Classification}

In order to compare active and inactive galaxies, the SDSS galaxies must be classified and the AGN
separated out.  \citet{BPT}, hereafter BPT, demonstrated that it is possible to distinguish Type~2
AGN from normal star-forming galaxies using the relative intensity ratios of easily-measured
emission lines. BPT found that the most useful intensity ratios were [OIII]/H$\beta$, [NII]/H$\alpha$,
[SII]/H$\alpha$ and [OI]/H$\alpha$, and plotting these ratios against each other yields diagnostic
diagrams.

\begin{figure}
\hspace{0.165in}
\includegraphics[height=3.2in,width=3.2in,angle=0]{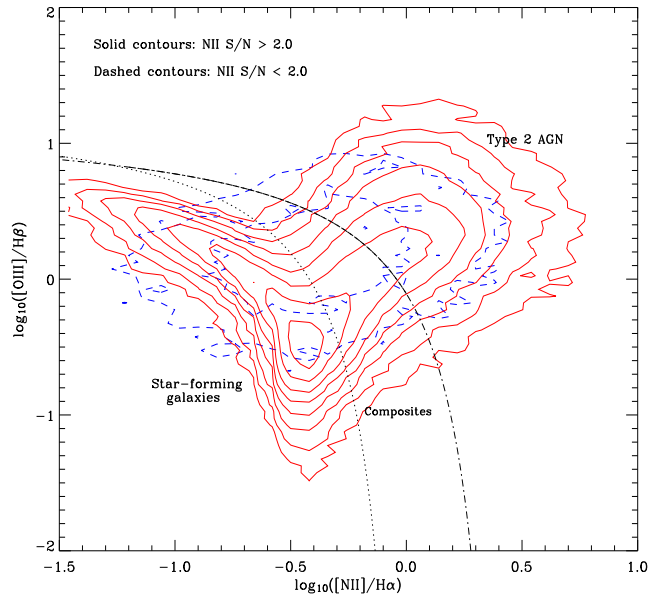}
\caption{BPT diagnostic diagram for DR4 SDSS galaxies, showing the distributions of the sample,
  above and below our $S/N$ cut of 2.0. The dot-dashed line is the Kew01 maximum star-burst line. The
  dotted line is the Kauf03 classification line.}
\label{fig:BPT}
\end{figure}

These emission lines have been accurately measured for all DR4 galaxies, as described by
\citealt{trem04}. The SDSS spectroscopic pipeline performs only a simple estimate of the stellar
continuum, which is good for strong lines, but not for weaker emission lines. The Tremonti et al.
pipeline on the other hand, estimates the stellar continuum by taking into account stellar Balmer
absorption, and hence recovers weaker nebular emission line features. This allows for a more
accurate classification scheme of SDSS galaxies.

We adopt a similar classification scheme to that devised by \citet{brinch04}, who split 
the sample into three main classes of galaxies based on their positions in the various BPT
diagrams. These classes were defined to be star-forming or passive galaxies, AGN, and composite
galaxies. Two extra sub-samples were derived for low signal-to-noise star-forming galaxies, and low
signal-to-noise AGN.  A final class was designated to ambiguous galaxies whose position in the
various BPT diagrams appeared to place them in more then one class - this class most likely to be
made up of galaxies with weak or no emission lines.

For our purpose, it is sufficient to separate `active' and `inactive' galaxies. We therefore split
the sample into three classes - star-forming/passive galaxies, Type~2 AGN, and composite
galaxies. We also use the \citealt{kew01a} (hereafter Kew01) maximum star-burst separation line
(dot-dashed line in Fig.~\ref{fig:BPT}) and the \citealt{kauf03b} (hereafter Kauf03) demarcation line
(dotted line in Fig.~\ref{fig:BPT}) to select our sample. Therefore, in this study, a galaxy is
defined to be a \textbf{Type~2 AGN} if
\begin{equation}
	\log([OIII]/H\beta) > 0.61/(\log([NII]/H\alpha)-0.05) + 1.3
\label{eqn:agn}
\end{equation}
or defined to be a purely \textbf{star-forming galaxy} if
\begin{equation}
	\log([OIII]/H\beta) < 0.61/(\log([NII]/H\alpha)-0.47) + 1.19.
\label{eqn:sf}
\end{equation}
where [OIII] refers to the [OIII] $\lambda$5007 emission line, and [NII] is the [NII] $\lambda$6584
line.\footnote{When calculating the logarithms in Eqns.~\ref{eqn:agn}
and~\ref{eqn:sf}, for each line flux, the value used is the higher of
the measured line flux and its 1-sigma uncertainty. This avoids
anomalously large, small or negative line ratios when one or both
lines are undetected.} Galaxies that fall in between these classification lines are galaxies that are classified as
AGN according to the more lenient Kauf03 separation line. These galaxies are kept as the separate
class of \textbf{composite galaxies}, as per Kauf03. 

We tested the sensitivity of our results to the galaxy classification, by slightly altering equations (1) and 
(2) in our selection process, making the composite sample smaller. The effects of such changes were minimal, 
so for simplicity, and comparison with other results, we continued to use the Kew01 and Kauf03 separation lines.

The Type~2 AGN and composite galaxies were also subjected to a signal-to-noise cut, to ensure
reliable AGN classification. We chose to cut at $S/N > 2.0$ in the [NII] $\lambda$6584 emission
line. As the [NII] emisson line flux for star-forming and passive galaxies can be low, no $S/N$ cut
was required for that classification. Figure~\ref{fig:BPT} shows a BPT plot for the overall DR4
sample. We have plotted the sample in two $S/N$ bins. The solid contours show the distribution of
galaxies with $S/N > 2.0$, while the dotted contours show the distribution of those with $S/N < 2.0$. The
latter distribution shows that these low $S/N$ galaxies are randomly distributed across the whole BPT
diagram, so will not accurately represent AGN.  Above a $S/N$ of two, the BPT diagram takes its
familiar form, so although this is a loose constraint it is still a reasonable cut for our statistical 
purposes in this paper, and allows us to keep the weaker AGN. A stricter cut of $S/N > 3.0$ was also 
tried, and found to have little impact on the overall results. Table~\ref{tab:number-summary} 
summarises the numbers of galaxies classified and rejected for our analysis.

Typical [OIII] luminosities in our active sample range from $10^5 L_{\sun}$ to $10^8 L_{\sun}$. The mean 
luminosity is approximately $10^6 L_{\sun}$: the majority of active galaxies would be considered 
to be `weak AGN' \citep{kauf03c} as less than 10\% of the active sample has $L[OIII] > 10^7 L_{\sun}$.

\begin{table}
\centering
\caption{Basic sample data following classification}
\begin{tabular}{lcr}
\hline
Subsample							&	Number		&	Percentage	\\
\hline
All galaxies ($0.01 < z < 0.16$)	&	359\,154	&	100.0		\\
SF + Passive						&	176\,596	&	49.2		\\
Type 2 AGN ($S/N > 2$ in [NII])		&	69\,783		&	19.4		\\
Composites ($S/N > 2$ in [NII])		&	64\,136		&	17.9		\\
Type 1 AGN (reject)					&	13\,331		&	3.7			\\
Unclassified (reject)				&	35\,308		&	9.8			\\
\hline
\end{tabular}
\label{tab:number-summary}
\end{table}

\begin{figure*}
\hspace{0.1in}
\includegraphics[height=6.5in,width=6.5in,angle=0]{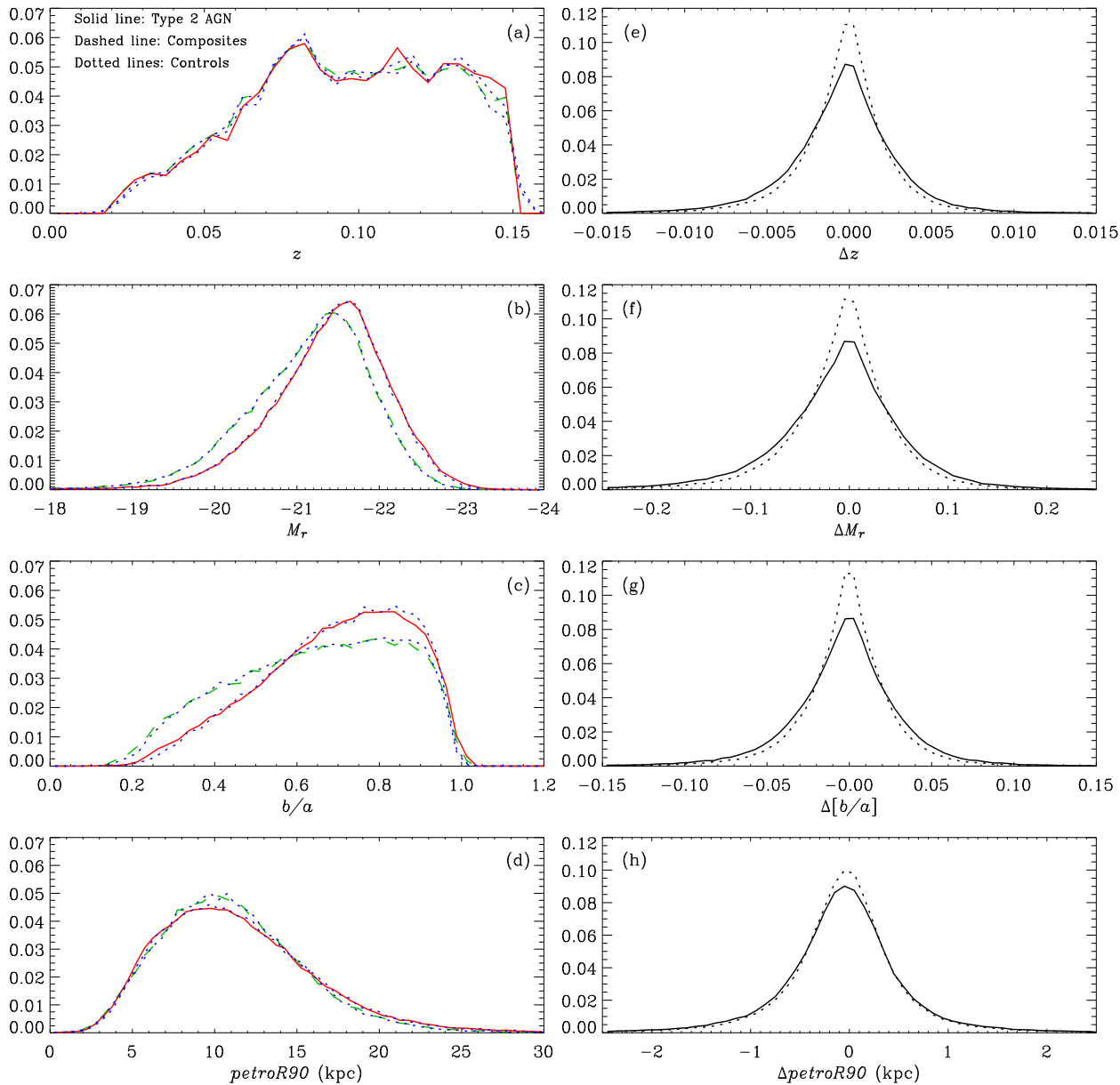}
\caption{Left-hand side: Normalised distributions of (a) redshift, (b) absolute magnitude, (c)
  aspect ratio, and (d) $petroR90$ radius, for our AGN (solid red lines) and composite
  galaxies (dashed green lines), along with their matched controls (dotted blue lines). Right-hand
  side: Normalised histograms of the individual differences between AGN/composites and each of their
  controls. The best controls are represented by the dotted lines, the 2nd best controls by the
  solid lines.}
\label{fig:matching}
\end{figure*}

\subsection{Control Galaxy Selection}

In order to perform a comparative study of active and inactive host galaxies, careful selection of
inactive control galaxies is required. Control matching is critical because the AGN classification
is strongly dependent on aperture effects, which can be minimised by matching redshift, and
inclination, which can be minimised by matching on aspect ratio. We also want to match to galaxies
of a similar size and luminosity. Therefore, for each of the active galaxies we selected two
inactive galaxies, closely matched in the following properties:
\begin{enumerate}
  \item Redshift, $z$, where $0.01 < z < 0.16$ (AGN and composites restricted
  to $0.02 < z < 0.15$);
  \item	Absolute $r$-band magnitude, $M_{r}$;
  \item	Aspect ratio, $b/a$, (Isophotal minor/major axis ratio in $r$-band);
  \item Radius, $petroR90$ (in arcsec), containing 90\% of the Petrosian flux, averaged
  over $r$- and $i$-bands.
\end{enumerate}

Isophotal major and minor axis values, given by $isoA$ and $isoB$ respectively,
are from the SDSS pipeline \textbf{\texttt{frames}}. \textbf{\texttt{frames}} measures the radius of
a particular isophote of the galaxy, as a function of angle, and then Fourier expands this
function. It then extracts $isoA$ and $isoB$ from the resulting coefficients. This
is done in all wavebands, but only the $r$-band values were used in this
analysis. $petroR90$ is also obtained through the \textbf{\texttt{frames}} pipeline. This
calculates the radius containing 90\% of the Petrosian flux for each band. At this stage
$petroR50$ is also derived, which is the radius containing 50\% of the Petrosian flux.

The $r$-band absolute magnitude used in this paper is given by
\begin{equation}
	M_{r} = r - k_{r} - 5 \log(D_{L} / 10{\rm\,pc})
\end{equation}
where $r$ is the Milky-Way-extinction-corrected Petrosian magnitude, $D_{L}$ is the
luminosity distance for a cosmology with $(\Omega_{m},\Omega_{\Lambda})_{0} = (0.3,0.7)$ and
$H_{0}$ = 70 km s$^{-1}$ Mpc$^{-1}$, and $k_{r}$ is the $k$-correction using the
method of \cite{blan03b}. \footnote{The $k$-corrections were derived from
\textbf{\texttt{kcorrect} v. 4.1.3.}}

Two controls were selected for an AGN by calculating the differences, $\Delta z$,
$\Delta [b/a]$, $\Delta M_{r}$, $\Delta petroR90$, between the AGN and
each control. The differences were then weighted, and summed as follows:
\begin{equation}
	\Delta C = \frac{| \Delta z |}{0.01} + \frac{| \Delta M_{r} |}{0.2} + \frac{| \Delta [b/a]
	|}{0.1} + \frac{| \Delta petroR90 |}{0.2\arcsec}
\end{equation}
The two controls with the lowest values of $\Delta C$ were then selected, and added to the
controls catalogue. This was done for all Type~2 AGN, and then done separately for all composite
galaxies. The weightings were such that we constrained the redshift and absolute magnitude more
tightly then the other two parameters [see Fig.~\ref{fig:matching}(e-h)].

\subsection{Sample Distributions}

The classification and control matching results in 2 pairs of galaxy catalogues: an AGN catalogue
of 64\,584 Type~2 AGN according to the Kew01 criteria (inclusive of LINERs), along with a catalogue
of 129\,168 control galaxies (Controls~A); and a catalogue of 60\,038 composite galaxies, with a
matched set of 120\,076 control galaxies (Controls~B).

\begin{figure*}
\hspace{0.2in}
\includegraphics[height=6.5in,width=3.0in,angle=90]{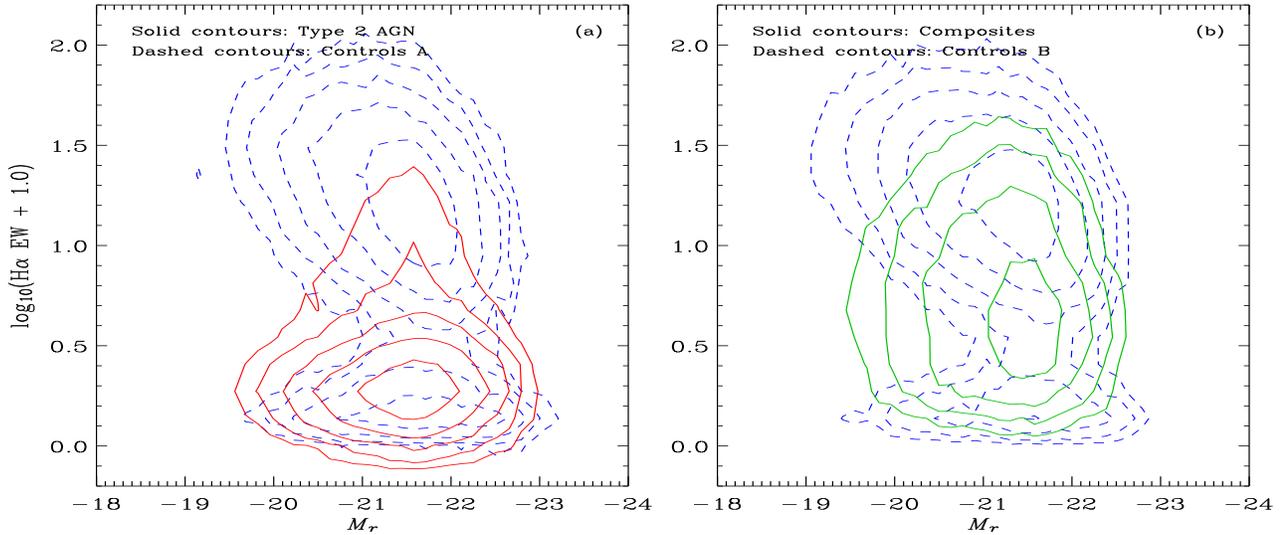}
\caption{Observed distribution of H$\alpha$ equivalent width vs.\ absolute magnitude. The contours
  are determined for galaxy number counts in 0.135 H$\alpha$ x 0.135 mag bins. The contour levels
  are on a logarithmic scale, starting at 64 and doubling each contour.  The solid lines in (a) are
  Type~2 AGN, and in (b) are composites.  The dashed lines are the control galaxies in both cases.}
\label{fig:H-alpha}
\end{figure*}

The resulting overall distributions of the selection properties are displayed in Fig.~\ref{fig:matching}(a-d), along
with the overall distributions of the differences in these properties between the AGN/composites and
each of their two controls - Fig.~\ref{fig:matching}(e-h). The best controls (dotted lines) result in a sharper peak
in the distribution, while the second best controls in our selection --- solid lines in Fig.~2(e-h)
--- show slightly broader wings. Control matching was done using all `inactive' galaxies each time,
so it is possible the same controls can be picked more than once, if they fit the selection criteria
for multiple AGN. It was not appropriate to make all control galaxies unique, as this would introduce 
systematic errors in the control matching.

\section{Results}

A significant feature in recent analyses of galaxy populations is the recognition that the
distribution of rest-frame colours is bimodal (e.g., \citealt{bald04}). There is a
well-characterised red sequence of predominantly massive, passively evolving early-type galaxies,
and a blue sequence of late type, star-forming galaxies. This is a key feature for understanding
galaxy evolution, and here we extend this analysis to identify where active galaxies fit into this
picture.

Past observations by, e.g., \cite{treu05}, suggest that galaxies transition from the blue to the red
sequence, in a manner consistent with ``cosmic downsizing'' \citep{cow96}; mergers between gas-rich
blue galaxies producing ellipticals, with a predicted phase of bright quasar activity which
terminates star-formation, and hence leads to the galaxy becoming red.

\subsection{Equivalent width--Magnitude relations}

The Balmer H$\alpha$ line is a clear indicator of active star formation \citep{kenni83}, and hence is a
good way of separating out the red and blue galaxy sequences \citep{bal04,haines07}. 
Fig.~\ref{fig:H-alpha}(a) shows the distribution of
the equivalent width (EW) of the H$\alpha$ lines in our AGN sample and the Controls~A sample, as a
function of absolute $r$-band magnitude.

The galaxy bimodality is clearly observed in the distribution of control galaxies, with late-type,
star-forming galaxies (blue sequence) occupying the top half of the plot, and early-type, passive
galaxies falling at low H$\alpha$ strength, and peaking at a slightly higher luminosity. The peak of
the AGN population falls along the blue edge of the red sequence, i.e., shows increased H$\alpha$ EW
relative to inactive red-sequence galaxies. The composite galaxies, shown in Fig.~\ref{fig:H-alpha}(b), however,
show the peak of their distribution is located in the valley between the blue and red sequences of
the Controls~B sample. Both these results suggest evidence for a ``transition'' population,
supporting galaxy evolutionary theories, in which star formation in massive galaxies ceases due to
the presence of an AGN, resulting in a migration from the blue sequence to the red sequence. (Note that 
there is no requirement that AGN or composites have significant $H\alpha$ emission, only NII $S/N > 2$ is required). 

\subsection{Colour-Magnitude Relations}

\begin{figure*}
\hspace{0.2in}
\includegraphics[height=6.5in,width=3.0in,angle=90]{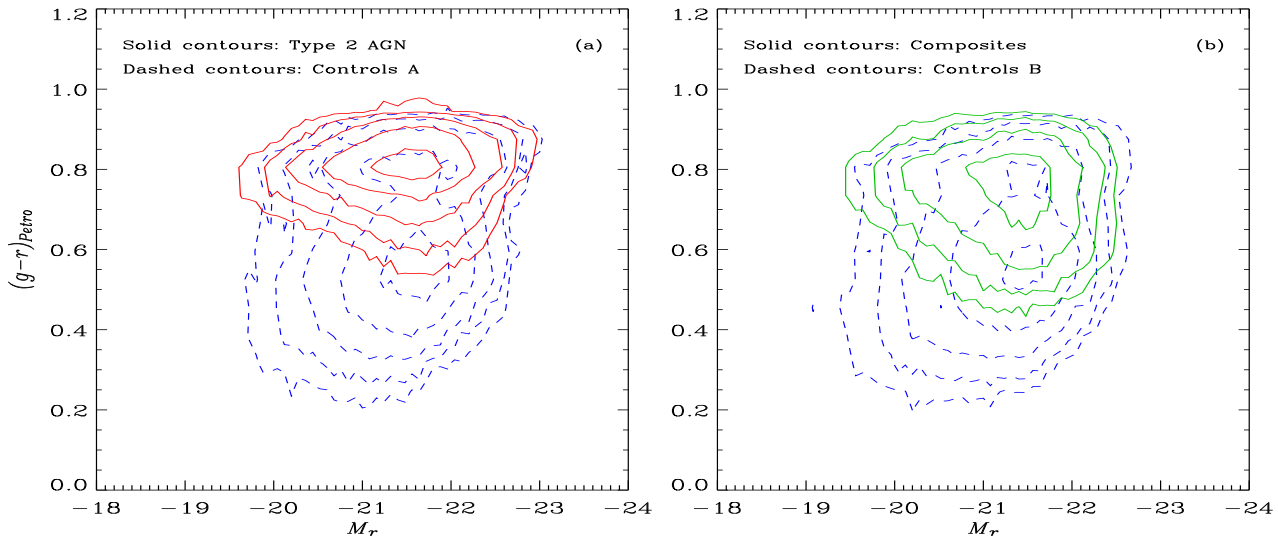}
\caption{Observed distribution of colour vs.\ absolute magnitude, for (a) Type~2 AGN, and (b)
  composite galaxies, and their respective controls. The contours are determined for galaxy number
  counts in 0.1 colour $\times$ 0.06 mag bins. The contour levels are on a logarithmic scale,
  starting at 16 and doubling each contour. The solid lines represent the `active' galaxies, while
  the dashed lines are the control galaxies.}
\label{fig:CMD}
\end{figure*}

$g-r$ colour reflects a longer timescale of star formation compared to H$\alpha$, which is
more representative of instantaneous star formation (or AGN-heated nebular emission). 
The colour-magnitude diagrams for our SDSS
sample are shown in Figure~\ref{fig:CMD}. The left-hand side plot again shows the Type~2 AGN and Controls~A, while
the right-hand panel shows the composite galaxies with Controls~B. The bimodality of colours can
be seen in the distributions of control galaxies in both cases. From the CMD in Fig.~\ref{fig:CMD}(a), the
majority of AGN now appear to reside along the peak of the red-sequence, rather than along its
blue-edge. The AGN population does, however, still extend to the red edge of the blue sequence as
well.

The location of the composite galaxies is less conclusive in this case [Fig.~\ref{fig:CMD}(b)]. The peak of this
population is not so clearly just in between the red and blue sequences, as was the case in the
H$\alpha$ plot, but rather shows a more extended, irregular distribution. The distribution tracks
the red sequence, in a manner similar to that of the AGN sample, but also extends to much bluer
colours, indicating ongoing star formation, but with AGN activity also taking
place. This result suggests that the composites do not necessarily represent a transition
population, as implied by Fig.~\ref{fig:H-alpha}(b).

\subsection{Colour-Concentration Relations as a function of $M_{r}$}

\begin{figure*}
\hspace{0.2in}
\includegraphics[height=4.0in,width=6.6in,angle=0]{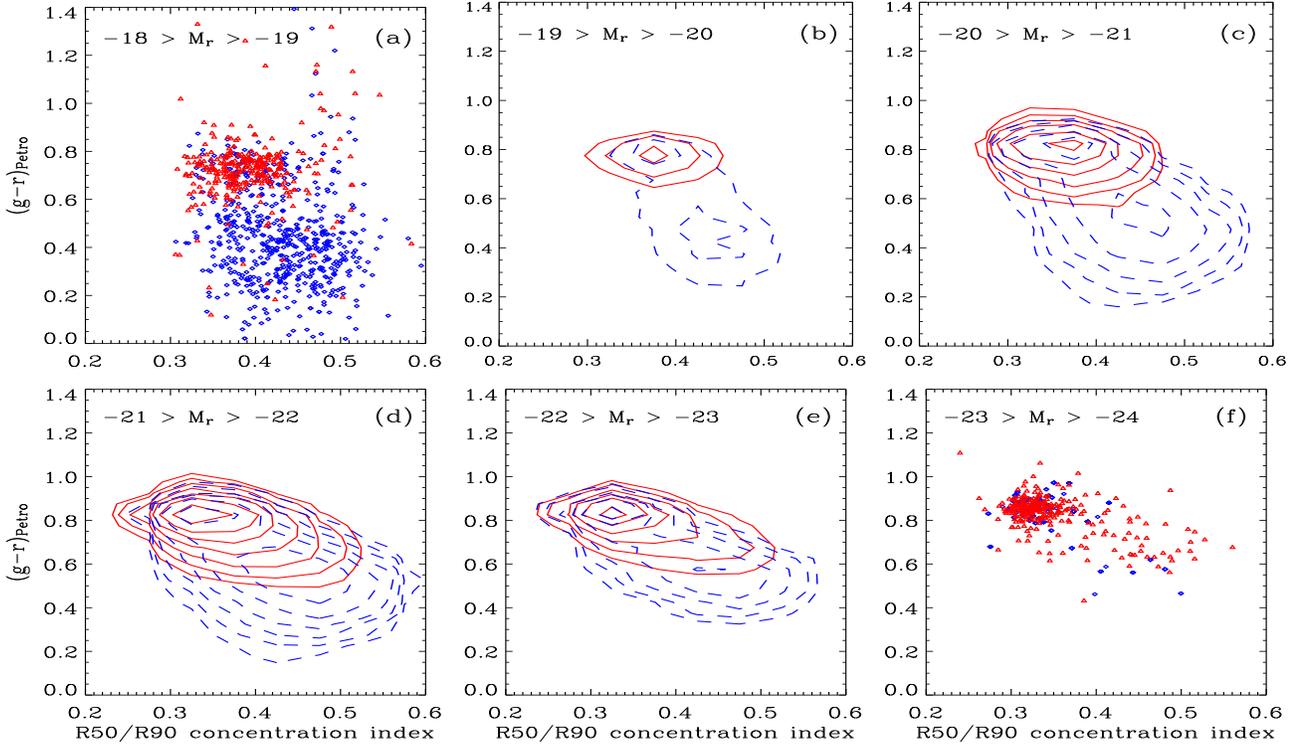}
\caption{Colour vs.\ concentration index for different luminosity ranges.  The contours are
  determined in 0.05 colour $\times$ 0.05 concentration bins.  The contour levels are on a logarithmic
  scale, starting at 64 and doubling each contour [Plots~(a) and~(f) had insufficient numbers for
  contours]. The solid red lines/red points are Type~2 AGN, and the dashed blue lines/blue points are `Controls~A'
  galaxies.}
\label{fig:colour-conc-agn}
\end{figure*}

Although we have demonstrated that the galaxy population separates into red and blue sequences, this
does not take into consideration the structure or morphology of galaxies. A robust method of
outlining the sequences is to consider a joint distribution in colour and structure
\citep{drive06}. Figure~\ref{fig:colour-conc-agn} shows the distribution of observed galaxies in colour versus concentration
index for six different absolute magnitude ranges.

The inverse concentration index, given by $C = petroR50/petroR90$, has been used in our
analysis. For typical galaxies, $C$ ranges from 0.3 (concentrated) to 0.55; for comparison, a
uniform disk would have $C = 0.75$.

In the distribution of the control galaxies (Controls~A; Figure~\ref{fig:colour-conc-agn}), the peak of the red sequence gets
slightly redder (0.75 to 0.90) and more concentrated (0.38 to 0.32) with brighter magnitude, while
the blue sequence also gets redder (0.45 to 0.60) but at an approximately constant concentration
index of 0.45. At fainter magnitudes ($-18 > M_{r} > -19$) there are few control
galaxies on the red sequence, and there is just a blue, low-concentration cloud. At brighter
magnitudes, it is the blue sequence that begins to disappear, with only the red sequence clearly
defined.  The AGN population peaks in approximately the same place as the peak of the red sequence
for the control sample, hence also gets redder and more centrally concentrated with brighter
magnitude.

\begin{figure*}
\hspace{0.2in}
\includegraphics[height=4.0in,width=6.6in,angle=0]{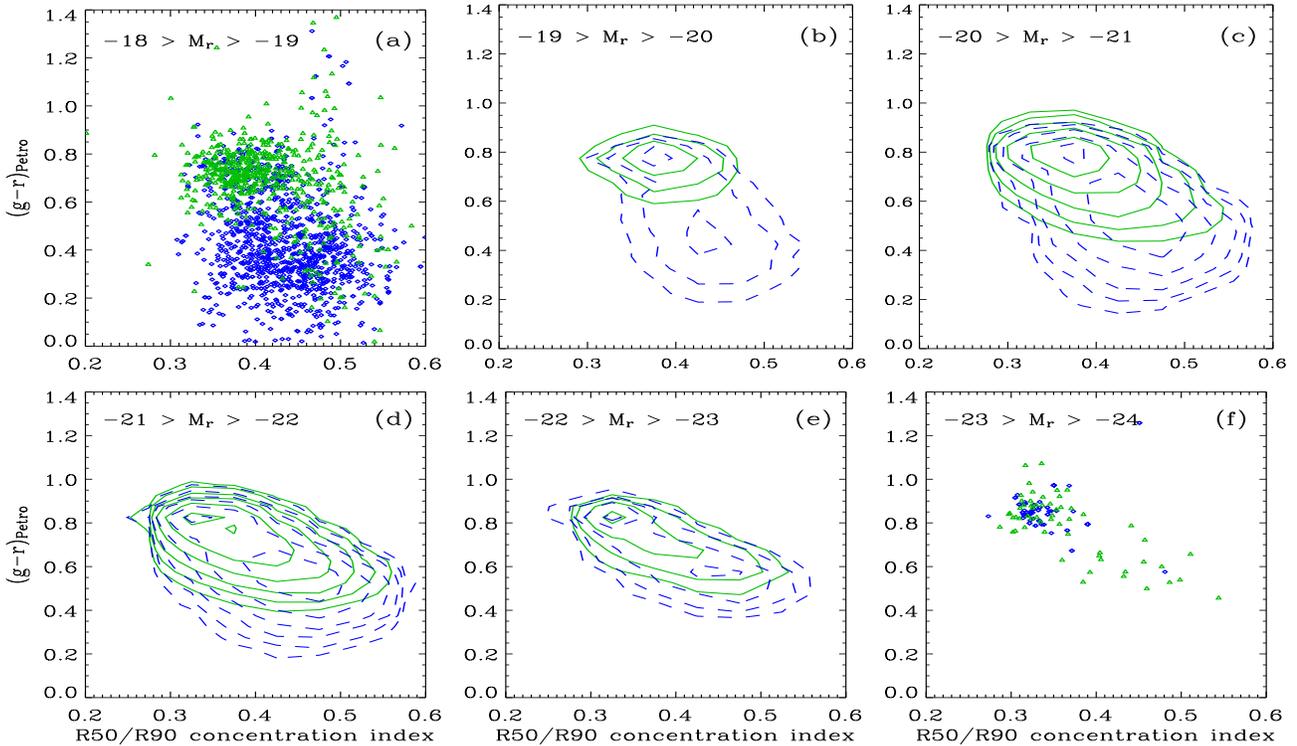}
\caption{Colour vs.\ concentration index for composite sample, in stellar-mass bins of 1
  $M_{r}$.  The solid green lines/green points are composites, and the dashed blue lines/blue points are
  `Controls~B' galaxies. See Fig.~\ref{fig:colour-conc-agn} for details.}
\label{fig:colour-conc-comp}
\end{figure*}

Figure~\ref{fig:colour-conc-comp} shows the colour-concentration relations for the sample of composite galaxies, along with
Controls~B. The distribution of Controls~B varies in the same way as Controls~A in Fig.~\ref{fig:colour-conc-agn}, but the
composite galaxies show a slight difference to the Type~2 AGN. The peak of the composite population
again becomes more centrally concentrated with increasing magnitude, but there is almost no change
in colour. A more complex morphology in the peak of the composite distribution also emerges in the
range $-20 > M_r > -23$ [Fig.~\ref{fig:colour-conc-comp}(c-e)], which is inconsistent with a simple transition population
implied by Fig.~\ref{fig:H-alpha}(b).

\subsection{Colour--Environment Relations as a function of $M_{r}$}

The relationship between galaxy colour and environment can also add constraints to the processes
involved in galaxy evolution. In this section we investigate the variation of the projected
neighbour density $\Sigma$, with $g-r$ colour and absolute magnitude.

Environmental densities were determined for SDSS galaxies by \citet{bald06}, using the projected
distance to the $N$th nearest neighbour that is a member of a density defining population
(DDP). The DDP were galaxies with $M_{r} < -20$. For each galaxy, $\Sigma$ was determined from the
4th and 5th nearest DDP galaxies within $\Delta zc = 1000 {\rm\,km\,s}^{-1}$. The environment
measurements were only reliable to $z<0.085$ so our sample sizes were reduced to 22\,747, 21\,666,
44\,592 and 42\,966, for Type~2 AGN, composites, Controls~A and Controls~B, respectively. This also
included a modest rejection of galaxies whose density measurements were highly uncertain.

\begin{figure*}
\hspace{0.2in}
\includegraphics[height=6.7in,width=5.1in,angle=90]{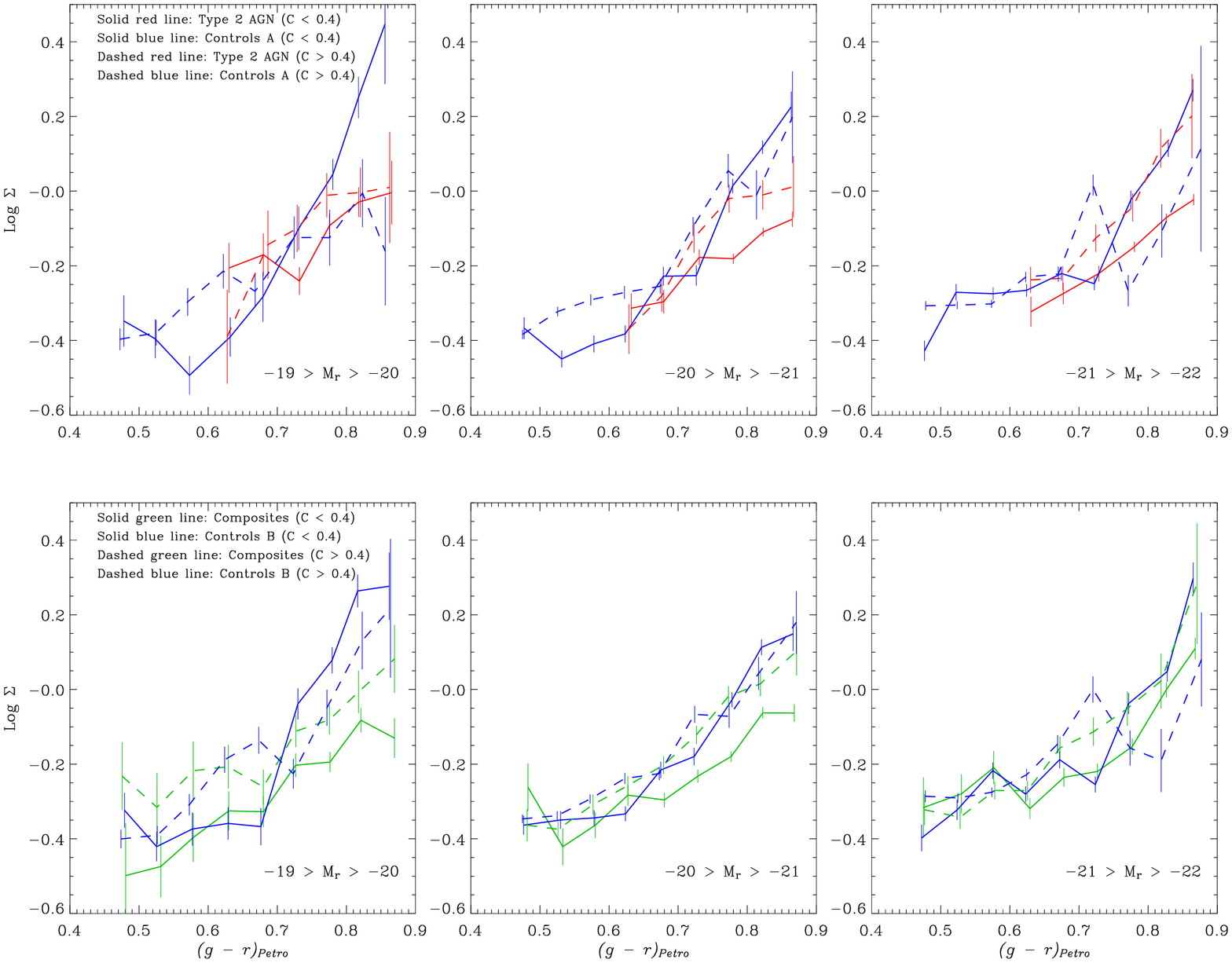}
\caption{Mean log $\Sigma$ versus $(g-r)_{Petro}$ colour, for three magnitude bins.  Top
  row: Type~2 AGN sample compared to Controls~A. Bottom row: Composites compared to Controls~B.  In
  both cases solid lines represent nominally bulge-dominated systems ($C < 0.4$), while
  dashed lines represent more disc-like systems ($C > 0.4$)}
\label{fig:mean-enviro}
\end{figure*}

Figure~\ref{fig:mean-enviro} shows how mean galaxy environment varies with colour and absolute magnitude. The top row
shows the AGN sample and Controls~A, while the bottom row shows the composites and Controls~B. We
have plotted the mean log $\Sigma$, in colour bins of 0.05, for three magnitude ranges. We have also
further split these up into two concentration bins; $C < 0.40$ (nominally bulge-dominated
galaxies; solid lines), and $C > 0.40$ (nominally disc galaxies; dashed lines).

From Fig.~\ref{fig:CMD}, a simple colour divide between the red and blue sequences occurs at
$(g-r)_{Petro} = 0.7$. Below 0.7 lie mainly late-type galaxies, and above 0.7 are the
early-type galaxies. Generally speaking then, Fig.~\ref{fig:mean-enviro} shows that late-type galaxies occur in less
dense environments, while an increase in colour across to the red sequence coincides with more dense
environments. Comparing just the solid lines in Fig.~\ref{fig:mean-enviro} --- i.e., bulge-dominated systems ---
suggests that on the red sequence, inactive galaxies occur in more dense environments than active
galaxies --- Type~2 AGN or composites. This effect is also stronger at fainter magnitudes. In
contrast, galaxies with $C > 0.4$ show little difference in environment density across the
different classes of galaxy, but all still show an increase in density with increasing colour.

At bluer colours, the error bars are greater as there are generally fewer late-type galaxies. Since
AGN are also predominantly red, there is little reliable data for this class at
$(g-r)_{Petro} < 0.6$.

\begin{figure}
\hspace{0.18in}
\includegraphics[height=4.7in,width=3.2in,angle=0]{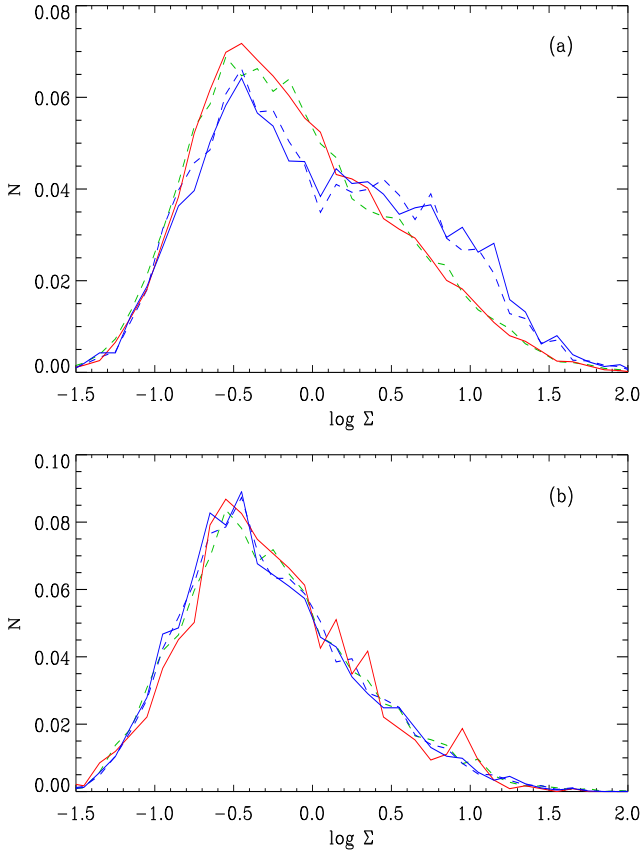}
\caption{Top panel: Normalised distribution of log $\Sigma$ for galaxies with
  $(g-r)_{Petro} > 0.7$ and $C < 0.4$ (red sequence).  Bottom panel: Normalised
  distribution of log $\Sigma$ for galaxies with $0.5 < (g-r)_{Petro} < 0.7$ and
  $C > 0.4$ (blue sequence).  Solid red lines are Type~2 AGN, and solid blue lines are
  Controls~A, while dashed green lines are composites, and dashed blue lines are Controls~B.}
\label{fig:distrib-enviro}
\end{figure}

Figure~\ref{fig:distrib-enviro} shows the distributions of $\log \Sigma$, for galaxies with 
$-19 > M_{r} > -22$. 
The top panel consists of galaxies with $(g-r)_{Petro} > 0.7$ and $C < 0.4$, so predominantly 
concentrated red galaxies, while the bottom panel shows the distribution for
galaxies with $0.5 < (g-r)_{Petro} < 0.7$ and $C > 0.4$. There is little or
no difference in the distribution of log $\Sigma$ across the various galaxy classes for the bluer,
disc galaxies (bottom panel), with all populations favouring less-dense environments, with a peak at log
$\Sigma = -0.5$.

For the concentrated red galaxies, however, this is not the case. As in Fig.~\ref{fig:mean-enviro}, we again see that
AGN hosts favour less dense environments, but this time we also see a slight bimodality in the
environments of inactive control galaxies. There is one peak, which falls in line with the peaks of
the AGN and composite populations, which is again at log $\Sigma = -0.5$, but there is also a
secondary peak at a higher density of approximately log $\Sigma = 0.75$, which most likely
corresponds to a unique population of galaxies in higher density environments.

\section{Discussion}

Having conducted a robust classification of `active' and `inactive' galaxies based on their emission
line properties and locations in the BPT diagram, we have created an AGN catalogue, with a carefully
selected control catalogue; and a catalogue of composite galaxies with another separate sample of
well-matched control galaxies. The control galaxy samples both display a bimodal distribution in
rest-frame colour, and also H$\alpha$ line strength. There is an AGN sequence that peaks along the
red sequence and extends down to the red edge of the blue sequence. Composite galaxies appear to be
slightly less luminous, and bluer, with the peak of the distribution possibly in the valley between
the red and blue sequences (see also \citealt{mill03}).

There are a number of possible explanations for this result: one interpretation is that star
formation maintains nuclear activity --- i.e., starburst winds are fuelling the central black hole
\citep{nand07}. However, the fact that the majority of our AGN lie on the red sequence,
suggests a shorter timescale for star formation associated with nuclear activity. This then points
in the direction of AGN activity suppressing star formation in massive galaxies
\citep{spring05,hopk05}. In this scenario, there needs to be an initial injection of gas to feed the
AGN, and then over time the AGN actually terminates star formation. In the local universe, AGN
feedback has been suggested as a mechanism to explain the observed effects of star formation
quenching (e.g., \citealt{bow06}). Gas expelled during the feedback is believed to cool and relax
the central regions of the galaxy resulting in passive evolution of the host galaxy, with stellar
winds fuelling the AGN. This theory could also account for the observed $M_{\bullet} - \sigma$
relation.

This would then naively predict that galaxies in the process of AGN feedback would lie in the
valley between the blue and red sequences, as they gradually reduce the amount of on-going star
formation until they end up on the red sequence. However, the efficiency of AGN feedback must be 
considered here, as it is unlikely that every AGN is able to suppress star formation in the galactic
bulge every time it ignites. It may be that feedback can only occur during a quasar phase, and not 
during periods of weak AGN activity. If this is the case, then the weak AGN would not be responsible 
for the galaxy appearing in a `transition' region, and the star-formation, AGN and dynamical 
timescales in the nucleus are likely to be the important factors for these galaxies.

While we do find many galaxies in this `valley',
particularly in the composite sample, it is difficult to conclude that this is a transitioning
population, because of the complex morphology of composites in the colour--concentration plane shown
in Fig.~\ref{fig:colour-conc-comp}. Composites appear to be either already on the red sequence, 
or on the red edge of the blue sequence with a distribution in concentration index that is consistent
with them remaining in the blue population. Other than stellar populations informing us whether or not 
there is a black hole, there is no obvious 
correlation between colour-concentration and the ability to fuel a black hole.

\cite{drory07} explain galaxy bimodality in terms of two different types of bulges --- classical
bulges, which are dynamically hot systems, and ``pseudobulges'', which are dynamically cold,
disc-like structures. The different structures are then thought to be indicators of the galaxy's
evolution.  Classical bulges form through violent relaxation during major mergers in early epochs,
when the environment density was high, making them red today. In contrast, pseudobulges are
considered disc-only galaxies that have not undergone a major merger since forming their disc, hence
are much younger, and bluer.  Fig.~\ref{fig:colour-conc-agn} supports this interpretation, and shows 
that AGN are generally classical bulge-dominated systems (E to Sa). This result, along with the 
$M_{\bullet} - \sigma$ relation indicates that the SMBH correlates with the classical bulge component, and not 
the galaxy as a whole, although the role of the host galaxy in triggering and fuelling the nuclear 
activity can not be ignored. 

Whilst Fig.~\ref{fig:colour-conc-comp} shows that the majority of composites are also bulge-dominated 
systems, there is also evidence for another population as well, which is again inconsistent 
with a transition population. However, making a typical blue-sequence, disk-dominated galaxy red, does 
not make it a `red-sequence' galaxy, as the latter are typically more massive, more concentrated, 
and of earlier Hubble-type, so it cannot be assumed that AGN activity in a less massive, late-type galaxy, 
is turning it into a red-sequence galaxy.

It has long since been established that early-type (red-sequence) galaxies are more clustered than
late-type (blue-sequence) galaxies \citep{dress80}, but in addition to this result, we now find that
active galaxies, despite being as red as inactive early-type galaxies, do not occur in such dense
environments as their inactive counterparts. In other words, the fraction of AGN amongst the
red-concentrated galaxies decreases in denser environments.  This result appears in contrast to
\citet{mill03} who found that the AGN fraction was constant with environment. However, the
combination of the passive-galaxy AGN fraction decreasing while the passive fraction increases could
maintain the constant AGN fraction over all galaxies. We also note that we use a significantly
larger sample with carefully matched controls.

\cite{li06} studied the clustering of narrow-line AGN in the SDSS, and on scales between 100\,kpc
and 1\,Mpc, conclude that AGN are preferentially located at the centres of dark matter halos. If
this is the case, then a proportion of the inactive red-sequence galaxies could be explained as
satellite galaxies in high density environments, which gives rise to the implied increased density
shown in Fig.~\ref{fig:mean-enviro}, and the secondary peak in Fig.~\ref{fig:distrib-enviro}(a).
Fuelling of AGN could be restricted in high-density environments except for galaxies at the centres
of dark matter halos.

\section{Conclusions}

We have carried out a robust classification of `active' and `inactive' galaxies in the SDSS, based
on their emission line properties and locations in the BPT diagram. We have compared the active
galaxies with well selected control galaxies matched in redshift, absolute magnitude, aspect ratio
and radius. We find that:

\begin{itemize}
	\item Type~2 AGN host galaxies occur mainly on the red sequence of the CMD, but show
	  increased levels of H$\alpha$ flux in their emission compared to inactive red-sequence
	  galaxies (Fig.~\ref{fig:H-alpha}(a) and Fig.~\ref{fig:CMD}(a));
	\item A separate class of composite galaxies appears to peak on the blue edge of the red
	  sequence on the CMD, whereas the peak of the H$\alpha$ distribution places composite
	  galaxies firmly in the valley between the blue and red sequences (Fig.~\ref{fig:H-alpha}(b) 
	  and Fig.~\ref{fig:CMD}(b));
	\item Colour-concentration relations, however, show a more complex, possibly double,
	  morphology in the peak of the composite distribution, rather than being in a valley 
	  (Fig.~\ref{fig:colour-conc-comp});
	\item AGN (and composites) are found in less dense environments on average then matched inactive
	  red-sequence galaxies. The more clustered inactive galaxies are likely to be satellite 
	  galaxies in high-density environments (Fig.~\ref{fig:mean-enviro} and Fig.~\ref{fig:distrib-enviro}).
\end{itemize}

The key to understanding this in more detail now lies in the dynamics of the central regions of
galaxies, to understand what activates some galaxies, but not others.

\section{Further Work}

The motivation for the construction of the parent sample presented here is an ongoing detailed study
of the 3D distribution and kinematics of gas and stars in active and inactive galaxies using the
IMACS-IFU on the Magellan telescope.
Equal numbers of low-redshift ($z < 0.05$), high-luminosity Type~1 Seyferts and Type~2 
Seyferts (which includes composites) are carefully selected by inspection of their SDSS spectra to 
ensure that they have strong emission lines, particularly in [OIII]. Control galaxies are then 
matched to the AGN in a similar way to the method presented in \S 3.2, but instead of two controls 
being produced, 10 to 15 potential matches are produced, and the best is selected following further 
inspection the SDSS images and spectra of the controls. This work is an extension of the SAURON-IFU 
study of nearby active and inactive galaxies by \cite{dumas07}.

The study of the characteristics of a large statistical sample (this paper)
complements detailed kinematic studies of a well-selected, but necessarily smaller sample of galaxies, 
which can identify the physical mechanisms that drive the trends observed in the statistical sample, 
through detailed, spatially resolved studies of individual galaxy nuclei.

\section*{Acknowledgments}

PBW acknowledges PPARC for a postgraduate studentship. CGM acknowledges financial support from the 
Royal Society. We thank Neil Nagar for observations and collaboration on the 3D survey, Phil James for 
useful discussions, and the anonymous referee for helpful comments that improved the paper.

We acknowledge NASA's Astrophysics Data System Bibliographic Services, the IDL Astronomy User's
Library, and IDL code maintained by David Schlegel (idlutils), and Michael Blanton (kcorrect) as
valuable resources. Data were taken from the SDSS catalogue provided by the MPA Garching Group.
Funding for the SDSS Archive has been provided by the Alfred P. Sloan Foundation, the 
Participating Institutions, the National Aeronautics and Space Administration, the National Science 
Foundation, the U.S. Department of Energy, the Japanese Monbukagakusho, and the Max Planck Society.


\label{lastpage}

\end{document}